\documentclass{iopart}
\usepackage[dvips]{graphicx}
\usepackage{amssymb}

\newcommand{\text}[1]{\hbox{\scriptsize\rm #1}}
\newcommand{\un}[1]{\mathrm{\,#1}}

\newcommand{\PRpt}{{\it Phys. Rep.}}
\newcommand{\AandA}{{\it Astron. Astrophys.}}
\newcommand{\published}[1]{\vspace{28pt plus 10pt minus 18pt}
     \noindent{\small\rm #1\par}}
\begin{document}
\title[Stochastic Background Search with ALLEGRO and LIGO]
{Stochastic Background Search Correlating ALLEGRO with
  LIGO Engineering Data}
\author{John T Whelan$^1$, Edward Daw$^2$,  Ik Siong Heng$^3$,
  Martin P McHugh$^1$ and Albert Lazzarini$^4$}
\address{$^1$ Department of Physics, Loyola University, New Orleans,
  Louisiana 70118, USA}
\address{$^2$ Department of Physics and Astronomy, Louisiana State
  University, Baton Rouge, Louisiana 70803, USA}
\address{$^3$ Max-Planck-Institut f\"{u}r Gravitationsphysik,
  Albert-Einstein-Institut, Aussenstelle Hannover, D-30167 Hannover,
  Germany}
\address{$^4$ LIGO Laboratory, California Institute of Technology,
  Pasdena, CA 91125, USA}
\begin{abstract}
  We describe the role of correlation measurements between the LIGO
  interferometer in Livingston, LA, and the ALLEGRO resonant bar
  detector in Baton Rouge, LA, in searches for a stochastic background
  of gravitational waves.  Such measurements provide a valuable
  complement to correlations between interferometers at the two LIGO
  sites, since they are sensitive in a different, higher, frequency
  band.  Additionally, the variable orientation of the ALLEGRO
  detector provides a means to distinguish gravitational wave
  correlations from correlated environmental noise.  We describe the
  analysis underway to set a limit on the strength of a stochastic
  background at frequencies near 900\,Hz using ALLEGRO data and data
  from LIGO's E7 Engineering Run.
\end{abstract}
\published{Published as {\CQG} \textbf{20,} S689 (2003);\\
  \copyright 2003 IOP Publishing Ltd, \texttt{http://www.iop.org/}}
\ead{jtwhelan@loyno.edu}
\maketitle

\section{Introduction}
\label{s:intro}

One of the gravitational wave (GW) sources targeted by the current
generation of ground-based interferometric and resonant detectors is a
stochastic background of gravitational waves (SBGW), produced by an
unresolved superposition of signals of astrophysical or cosmological
origin\cite{Christensen:1992,Allen:1997,Maggiore:2000}.  Direct
observational limits
can be set on a SBGW by looking
for correlations in the outputs of a pair of GW detectors.  This has
been done using two resonant-bar detectors \cite{Astone:1999}, two
``prototype'' interferometers \cite{Compton:1994}, and is
currently being done with the two kilometer-scale LIGO
interferometers\cite{Allen:2002,StochS1} (IFOs).  In this paper we describe the
first known effort to set a limit on a SBGW with correlations between
an IFO and a bar.  This pair of detectors--the LIGO Livingston
Observatory (LLO)\cite{LLO} in Livingston, LA and the ALLEGRO resonant
bar detector\cite{ALLEGRO} in Baton Rouge, LA--is separated by only
40\,km, the closest
pair among the ten modern ground-based GW detector sites. \cite{GWIC}
This makes
it an attractive pair for probing the stochastic GW spectrum around
900\,Hz.  In addition, the ALLEGRO bar can be rotated, changing the
response of the correlated data streams to stochastic GWs and thus
providing a means to distinguish correlations due to a SBGW from those
due to correlated environmental noise.

In Sec.~\ref{s:stoch} we review the standard measure of stochastic
background strength and the cross-correlation technique used to search
for a SBGW.  In Sec.~\ref{s:prev} we describe the previous limits
set on a SBGW with ground-based GW detectors.  In
Sec.~\ref{s:allegro} we describe the key features of the
LLO-ALLEGRO correlation experiment in general.  Section~\ref{s:e7}
describes the data taken during LIGO's E7 engineering run and the
analysis currently underway using those data.  Finally,
Sec.~\ref{s:future} describes the prospects for future correlation
experiments using ALLEGRO and LIGO science data.

\section{Stochastic Background Measurements}
\label{s:stoch}

A SBGW is assumed for simplicity to be  isotropic, unpolarized, Gaussian,
and stationary.  Subject to these assumptions, the stochastic
GW background is completely described by its power
spectrum.  It is conventional to express this spectrum in terms of the
GW contribution to the cosmological parameter
$\Omega=\rho/\rho_{\text{crit}}$:
\begin{equation}
  \label{eq:omegagw}
  \Omega_{\text{GW}}(f)=\frac{1}{\rho_{\text{crit}}}
  \frac{d\rho_{\text{GW}}}{d\ln f}=\frac{f}{\rho_{\text{crit}}}
  \frac{d\rho_{\text{GW}}}{df}
  \ .
\end{equation}
Note that $\Omega_{\text{GW}}(f)$ has been constructed to be
dimensionless, and represents the contribution to the overall
$\Omega_{\text{GW}}$ per \emph{logarithmic} frequency interval.  In
particular, it is \emph{not} equivalent to $d\Omega_{\text{GW}}/df$.
Note also that since the critical density $\rho_{\text{crit}}$, which
is used in the normalization of $\Omega_{\text{GW}}(f)$, is
proportional to the square of the Hubble constant $H_0$ \cite{Kolb:1990},
it is convenient to work with $h_{100}^2\Omega_{\text{GW}}(f)$, which
is independent of the observationally determined value of
$h_{100}=\frac{H_0}{100 \un{km}/\un{s}/\un{Mpc}}$.

The standard method to search for such a background is to
cross-correlate the outputs of two gravitational wave detectors
\cite{Christensen:1992}.  If the noise in the two detectors is
uncorrelated, the only non-zero contribution to the average
cross-correlation will come from the stochastic GW background.  In the
optimally-filtered cross-correlation method (described in more detail
in \cite{Allen:1997,Allen:1999,Whelan:2001}), one calculates a
cross-correlation statistic
\begin{equation}
  Y
  =
  \int dt_1\, dt_2\, {h_1(t_1)}\, {Q(t_1-t_2)}
  {h_2(t_2)}
  \label{eq:CCstat}
  =
  \int df\,{\widetilde{h}_1^*(f)}\, {\widetilde{Q}(f)}\,
  {\widetilde{h}_2(f)}
\end{equation}                                
where $h_{1,2}(t)$ are the data streams from the two detectors,
$\widetilde{h}_{1,2}(f)$ are their Fourier transforms, and $Q(t_1-t_2)$ (with
Fourier transform $\widetilde{Q}(f)$) is a suitably-chosen optimal
filter.  It is sensitive\cite{Allen:1999} to backgrounds on the
order of
\begin{equation}
\label{eq:CCsens}
\Omega^{\text{UL}} 
\sim
\left(
  {T}
  \int df\frac{
    {\gamma_{12}^2(f)}}{f^6 {P_1(f) P_2(f)}}
  \right)^{-1/2}
  \ .
\end{equation}
This sensitivity improves with time and is limited by the power
spectral densities $P_{1,2}(f)$ of the noise in the two detectors.
The factor
\begin{equation}
  \gamma_{12}(f)={d_{1ab}}\ {d_2^{cd}}\
  {{5\over 4\pi}\int_{S^2}}{d^2\hat\Omega}\
  {e^{i2\pi f{\hat\Omega}{\cdot}
      {\Delta \vec x}{/c}}}\
  {P^{ab}_{cd}(\hat\Omega)}
\end{equation}
in the numerator of the integral is the \emph{overlap reduction
  function} \cite{Flanagan:1993}, which describes the observing
geometry.  Here $P^{ab}_{cd}(\hat\Omega)$ is a projector onto
symmetric traceless tensors transverse to a direction $\hat{\Omega}$
and $d_{1,2}^{ab}$ are the \textit{detector response tensors} for the
two detectors.  These are the tensors with which the metric
perturbation $h_{ab}$ at the detector should be contracted to obtain
the \textit{gravitational wave strain} $h=d^{ab}h_{ab}$.  If $u_a$ and
$v_a$ are unit vectors pointing in the directions of an
IFO's two arms, its response tensor is
\begin{equation}
  d_{ab} = \frac{1}{2}(u_a u_b - v_a v_b)
\end{equation}
while the response tensor for a resonant bar whose long axis is
parallel to the unit vector $w_a$ is
\begin{equation}
  d_{ab} = w_a w_b
  \ .
\end{equation}

\begin{figure}[htbp]
  \vspace{5pt}
  \begin{center}
    \includegraphics[height=3.5in]{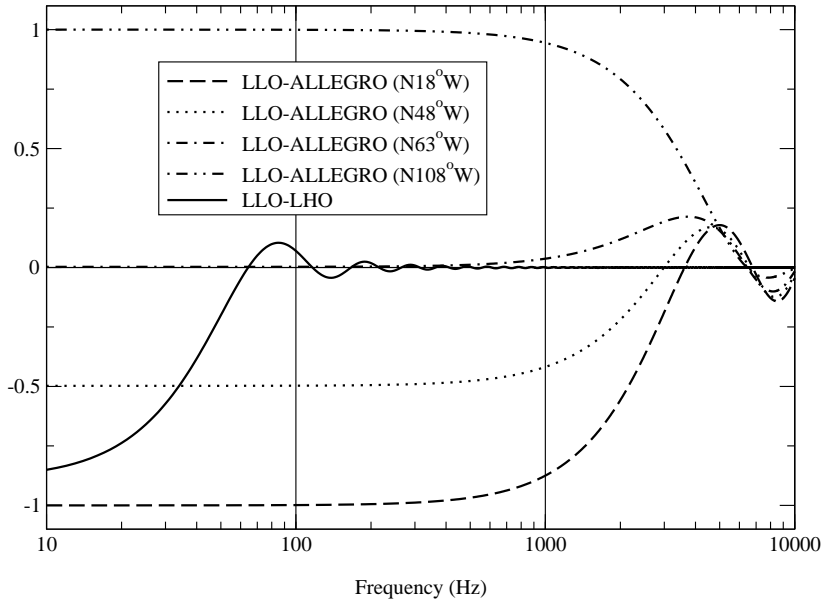}    
  \end{center}
  \caption{
    The overlap reduction function for LIGO Livingston Observatory
    (LLO) with ALLEGRO and with LIGO Hanford Observatory (LHO).  The
    four LLO-ALLEGRO curves correspond to four orientations:
    ``N$18^\circ$W'' is nearly parallel to the y-arm of LLO
    (``anti-aligned''); ``N$108^\circ$W'' is nearly parallel to the
    x-arm of LLO (``aligned''); ``N$63^\circ$W'' is halfway in between
    these two orientations (a ``null alignment'' midway between the
    two LLO arms); ``N$48^\circ$W'' is an intermediate alignment in
    which ALLEGRO was also operated during the E7 run.
  }
  \label{fig:overlap}
\end{figure}

The overlap reduction function is equal to unity for the case of a
pair of IFOs (or an IFO and a bar) at the same
location with their arms aligned, and is suppressed as the detectors
are rotated out of alignment or separated from one another.  The
frequency dependence comes about for the following reason: if the
wavelength is comparable to or smaller than the separation
between two detectors, the detectors will see different phases of the
wave at the same time, and this phase difference will depend on the
direction of propagation of the wave.  Since the stochastic GW
background is assumed to be isotropic, averaging over different
propagation directions suppresses the sensitivity of a pair of
detectors to high-frequency waves.  For example, a wave whose
wavelength is twice the distance between the two detectors will drive
them $180^\circ$ out of phase if it travels along the line separating
them, but \emph{in} phase if its direction of propagation is
perpendicular to this line.  Figure~\ref{fig:overlap} shows the
overlap reduction functions for several detector pairs of interest.

\section{Previous Results}
\label{s:prev}

The current best upper limit on a SBGW from direct observation with GW
detectors is $h_{100}^2\Omega_{\text{GW}}(900\un{Hz})\le 60$
\cite{Astone:1999}, set by correlating the resonant bar detectors
Explorer\cite{Explorer} (in Geneva, Switzerland) and
Nautilus\cite{Nautilus} (near Rome, Italy).\footnote{The same group
  had previously\cite{Astone:1996} set a limit of
  $h_{100}^2\Omega_{\text{GW}}(900\un{Hz})\le 100$ using a single bar
  detector.}  A broad-band limit of $h_{100}^2\Omega_{\text{GW}}(f)\le
3\times 10^5$ was set using a pair of ``prototype interferometers''
\cite{Compton:1994}, and more recently an analysis of E7 engineering
data from the LLO and LHO sites \cite{Allen:2002} set a limit of
$h_{100}^2\Omega_{\text{GW}}(40\hbox{--}215\un{Hz})\le 7.7\times
10^4$.  Analysis \cite{StochS1} of LLO and LHO data from LIGO's first
science run (S1) is expected to improve substantially upon this limit.

More stringent upper limits can be set on astrophysical grounds.  They
are detailed elsewhere \cite{Allen:1997,Maggiore:2000,StochS1}, but we
mention the
bound from big-bang nu\-cle\-o\-syn\-the\-sis\cite{Kolb:1990,Maggiore:2000},
which states that a cosmological SBGW is limited by
\begin{equation}
  \int_{10^{-8}\un{Hz}}^\infty \frac{df}{f}h_{100}^2\Omega_{\text{GW}}(f) 
  \le 10^{-5}
  \ .
\end{equation}
This broad-band limit tells us that any cosmologically interesting SBGW
almost certainly lies several orders of magnitude below the existing
limits.

\section{ALLEGRO-LLO Correlations}
\label{s:allegro}

Due to the 3000\,km distance between the two LIGO detectors, the
overlap reduction function (see Fig.~\ref{fig:overlap}) limits the
range of frequencies at which they are sensitive to a stochastic
background.  The separation distance is half of a GW wavelength at a
frequency of 50\,Hz, so the upper end of the frequency range is a few
times this.  (The lower end of the frequency range is set by the
seismic noise in the two detectors.)
The ALLEGRO bar detector is far closer to the LIGO Livingston site
than the LIGO Hanford site is, with only about 40\,km separating the
two and a ``half-wavelength'' frequency of 3750\,Hz.  Thus the
observing geometry for ALLEGRO-LLO allows for observations of
correlations out to much
higher frequencies.  On the other hand, the sensitivity of ALLEGRO
(see Fig.~\ref{fig:allegronoise}) is
concentrated in two narrow frequency bands in the vicinity of 900\,Hz,
so correlations between ALLEGRO and LIGO Livingston probe a different
part of the frequency domain than correlations between the two LIGO
detectors. The major challenge in detecting a SBGW is accounting for the
effects of noise which may be correlated between the two detectors.
Even in widely separated detectors such as LLO and LHO there may exist
correlations (the most obvious example being the due to the power line
frequency and its harmonics) which could mimic or mask a real SBGW.
With detectors in much closer proximity,
such as ALLEGRO and LLO, the problem could potentially be more severe.
A method proposed by Lazzarini and Finn \cite{Finn:2001} is to rotate
the ALLEGRO bar about a vertical axis and measure the
cross-correlation for different alignments of the ALLEGRO-LLO pair.
Changing the alignment of the bar from one of the interferometer arms
to the other changes the sign of the expected correlation due to a
SBGW. Thus any correlated noise which is independent of alignment can
be removed. Another possibility is to record data with the bar aligned at
$45^\circ$ from the arms where no correlation due to a SBGW is
expected. This ``off source''
measurement gives an uncontaminated estimate of the cross-correlated noise.

\subsection{{LLO-ALLEGRO Co\"{\i}ncidence Operation}}

\begin{figure}[htbp]
  \centering
  \includegraphics[height=3.5in]{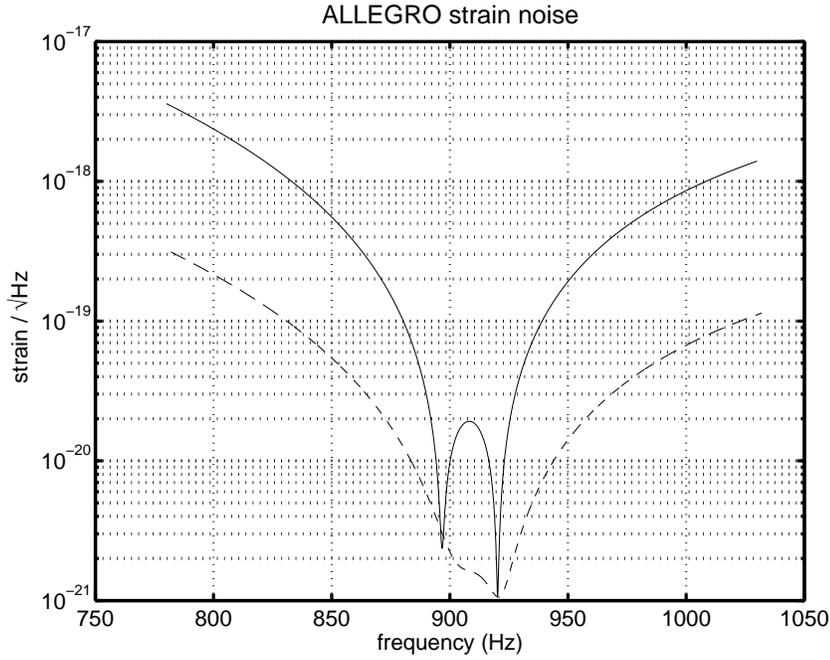}
  \caption{The figure shows the ALLEGRO strain noise curve for
    operation during E7 (solid) \protect\cite{Mauceli:1996} as well as
    the projected improvement
    for operation with the new transducer during S2
    (dashed)\protect\cite{Harry:1999}}
  \label{fig:allegronoise}
\end{figure}

The ALLEGRO detector took data during LIGO ``E7'' Engineering Run
that took place from December 2001 to January 2002. In the following
section
we discuss the initial investigations underway. Following the the E7
run ALLEGRO went off-line for a major upgrade. Beginning in the summer
of 2002 a new
transducer \cite{Harry:1999} was installed which should
substantially improve the sensitive bandwidth of the bar detector.
Figure~\ref{fig:allegronoise} shows the projected improvement in
bandwidth and sensitivity. The installation was ongoing during the
time of the LIGO S1 science run in August-September 2002.  Following
the installation, a series of cool-down attempts were made. A
succession of repairs were done on the vacuum system and a successful
cool-down was finally made starting in February 2003. This allowed
data to be taken with ALLEGRO during approximately
the second half of the LIGO S2 run.

\section{Investigations with LIGO Engineering Data}
\label{s:e7}

From December 28, 2001 to January 14, 2002, LIGO held its seventh
Engineering Run (E7), the last engineering run before the start of
scientific operation in August 2002.  The ALLEGRO detector was in
operation for the end of that run.  Table~\ref{tab:E7ALLEGRO}
summarizes the ALLEGRO data which overlap with science-quality LLO
data.
\begin{table}[htbp]
  \begin{center}
    \begin{tabular}{|r|r|r|}
      \hline
      Azimuth & $\gamma(907\un{Hz})$ &  Clean LLO Segments \\
      \hline
      \hline
      {N$48^\circ$W} & {-0.43}
      & {$134 \times 15\un{min}=$33:30} \\
      \hline
      {N$18^\circ$W} & {-0.90}
      & {$82 \times 15\un{min}=$20:30} \\
      \hline
      {N$63^\circ$W} & {-0.03}
      & {$110 \times 15\un{min}=$27:30} \\
      \hline
    \end{tabular}
  \end{center}
  \caption{Summary of ALLEGRO data co\"{\i}ncident with LLO operation
    during E7.  The table lists ALLEGRO's alignment in degrees West of
    North, the resulting value of the overlap reduction function
    $\gamma(f)$ (see Fig.~\protect\ref{fig:overlap}) in 
    ALLEGRO's sensitivity band, and the number of
    15-minute segments of clean LLO data which co\"{\i}ncide with
    ALLEGRO data in each alignment.
  }
  \label{tab:E7ALLEGRO}
\end{table}

If no excess correlations are found, and we assume that all of the
ALLEGRO data have a level of noise given by the solid curve in 
Fig.~\ref{fig:allegronoise}, and all of the
LLO data are of a quality described by \cite{E7StrainComp}, (\ref{eq:CCsens})
tells us to expect to set an upper limit of around $1.2\times
10^4$.\footnote{This would actually be better than the limit of
  $7.7\times10^4$ set in \cite{Allen:2002}, which is understandable
  because it involves only one engineering-quality LIGO data stream,
  while the LLO-LHO analysis involves two.  The sensitivity of at both
  LIGO sites improved markedly between E7 and the science runs.
  \cite{E7toS1StrainLLO}}

Analysis of the data described in Table~\ref{tab:E7ALLEGRO} is
currently in progress.  In the remainder of this section, we outline
some of the technical challenges involved in this analysis.

First, the nature of the data recorded is different: LIGO data
consists of a real time series $h_L(t)$, sampled at 16384\,Hz, which
is related
by a linear transfer function to the gravitational wave strain in the
detector.  The Fourier transform $\tilde{h}_L(f)$ of this data stream contains
frequencies from $-8912$\,Hz to $8912$\,Hz, although the negative
frequency components are all determined by the positive frequency
ones.  Because the ALLEGRO detector's sensitivity is confined to a band
with $\Delta f<f$, its data stream $h_A(t)$ is heterodyned (mixed with
an oscillating signal at frequency $f_H=907$\,Hz) and low-pass
filtered (before it is
digitized) and then sampled at the reduced rate of 250\,Hz.  The
result is to produce a complex signal
$h_A^H(t)=e^{i2\pi(-f_Ht+\phi_A)}h_A(t)$ whose Fourier transform
$\tilde{h}_A^H(f)\approx\tilde{h}_A(f_H+f)e^{i2\pi\phi_A}$ contains
completely independent data at frequencies
$-125\,\mathrm{Hz}<f<125\,\mathrm{Hz}$, which are approximations of
the Fourier transforms of the
data in the original time series between $(907-125)$\,Hz and
$(907+125)$\,Hz.

The initial approach we are taking to correlating these data streams
with different sampling frequencies is to heterodyne and resample the
LIGO data in software (producing an
$h_L^H(t)=e^{i2\pi(-f_Ht+\phi_L)}h_L(t)$ whose Fourier transform is
$\tilde{h}_L^H(f)\approx\tilde{h}_L(f_H+f)e^{i2\pi\phi_L}$) and then
correlate it with the ALLEGRO data as
though they were any two heterodyned data streams.  This requires a
certain amount of care, because the phase $\phi_L$ at the start of a
stretch of
data of the reference oscillator for the LIGO heterodyning is
specified (because it's done in software), while the phase $\phi_A$ of
the
ALLEGRO reference oscillator is only known if it is experimentally
determined.  In principle, it is not difficult to keep track of this,
but it is also possible to post-process the cross-correlation
statistics to account for ignorance of the phase difference
$\phi_A-\phi_L$ between the two
oscillators.\cite{McHugh:2002}

Additional technical challenges include converting ALLEGRO data into
the frame format used by LIGO\cite{frame} so that it can be processed
by the LIGO Data Analysis System\cite{ldas}.  Fortunately, the LIGO
stochastic search codes\cite{Whelan:2001,lal,LALWrapper} have been
designed
with an eye to working on both heterodyned and non-heterodyned data,
and the data structures describing site geometry\footnote{LAL
  Software Documentation, version 2.0, Section
  9.2; available from \cite{lal}}
allow for either
interferometric or resonant bar detector geometries.

\section{Future Outlook}
\label{s:future}

The upper limit we expect to set from E7 data is several orders of
magnitude above the current direct observational
limit\cite{Astone:1999} at the frequencies in question.  But the
practical application of the LLO-ALLEGRO cross-correlation method to
engineering data will pave the way for future such observations, whose
sensitivity is expected to improve for several reasons:
\begin{enumerate}
\item Most substantially, the sensitivity of the LLO IFO has already
  improved markedly\cite{E7toS1StrainLLO} and will continue to improve as the
  instrument approaches its design sensitivity\cite{E7StrainComp}.
\item ALLEGRO's new transducer will increase the bandwidth of the bar
  detector (see the dashed curve in Fig.~\ref{fig:allegronoise}).
\item Future observing runs will provide more co\"{\i}ncident data
  than the 80 hours or so taken during E7.
\end{enumerate}
Assuming a full year of co\"{\i}ncident data in optimal alignment,
with LLO operating at design sensitivity and ALLEGRO operating at the
sensitivity given by the dashed curve in
Fig.~\ref{fig:allegronoise}, we estimate using (\ref{eq:CCsens})
that an upper limit could be set in the range $890<f<930$ of
$h_{100}^2\Omega_{\text{GW}}(f)\lesssim 10^{-2}$.

\ack

We would like to thank everyone at the LIGO and ALLEGRO projects and
especially the LSC's stochastic sources upper limits group.  JTW and
ISH gratefully acknowledge the University of Texas at Brownsville and
Louisiana State University, respectively, where this project was
begun.
This work was supported by the National Science Foundation under
grants PHY-9981795 (UTB) and PHY-9970742 (LSU), and by NASA contract
JPL1219731 (UTB).
Figure~\ref{fig:overlap} was made using the LALApps program library,
based on the LAL software library\cite{lal}.

\end{document}